\documentstyle[amssymb,aps,prl,epsfig]{revtex}

\begin{document}
\draft

\title{Interacting electrons in a nearly straight
quantum wire}
\author{T. Rejec$^1$, A. Ram{\v s}ak$^{1,2}$, and J.H. Jefferson$^3$}
\address{$^1$J. Stefan Institute, SI-1000 Ljubljana, Slovenia\\
$^2$Faculty of Mathematics and Physics, University of Ljubljana,
SI-1000 Ljubljana, Slovenia\\
$^3$DERA, St. Andrews Road, Great Malvern, Worcestershire WR14 3PS,
England}
\date{July 26, 2000}
\maketitle

\begin{abstract}
\widetext
\smallskip
We study the conductance threshold of a clean nearly straight quantum wire
in which a single electron is bound. This exhibits spin-dependent
conductance anomalies on the rising edge to the first conductance plateau,
near $G=0.25(2e^{2}/h)$ and $G=0.7(2e^{2}/h),$ related to a singlet and
triplet resonances respectively. We show that the problem may be mapped on
to an Anderson-type of Hamiltonian and calculate the energy dependence of
the energy parameters in the resulting model.
\end{abstract}
\vskip 1cm

Conductance steps in various types of quantum wire have now been
observed, following the pioneering work in Refs.~\cite{Rejec_wees88}.
These first experiments were performed on gated two-dimensional
electron gas (2DEG) structures, though similar behaviour has now
been observed in other quantum wire structures \cite{Rejec_wires}.
Whilst these experiments strongly support the idea of ballistic
conductance in quantum wires, and are in surprising agreement
with the now standard Landauer-B\"{u}ttiker formalism \cite
{Rejec_landauer57}, certain anomalies can arise which are
spin-dependent and are believed to originate from
electron-electron interactions. In particular, already in early
experiments a structure is seen in the rising edge of the
conductance curve \cite{Rejec_wees88}, starting at around
$0.7(2e^{2}/h)$ and merging with the first conductance plateau
with increasing energy. Later experiments also clearly showed
anomalies near $G=0.25(2e^{2}/h)$ \cite {Rejec_wires}. Recently we
have shown that these conductance anomalies are consistent with
an electron being weakly bound in wires of circular and
rectangular cross-section, giving rise to spin-dependent
scattering resonances \cite{Rejec_rejec99}.

In this paper we present further results on the above anomalies,
related to singlet and triplet resonances for a propagating
electron at the Fermi energy scattering from the weakly bound
electron. This two-electron problem (solved exactly) is then
mapped onto a many-electron Anderson-type model for which the
most important matrix elements are retained.

We consider quantum wires which are almost perfect but for which
there is a very weak effective potential, giving rise to a bound
state. Such an effective potential can arise, for example, from a
smooth potential due to remote gates or a slight buldge in an
otherwise perfect wire. We consider this latter situation for the
cases of quantum wires with circular cross-section, appropriate
for, e.g., `hard-confined' v-groove wires. The cross-sections of
these wires are sufficiently small that the lowest transverse
channel approximation is adequate for the energy and temperature
range of interest. The smooth variation in cross-section also
guarantees that inter-channel mixing is negligible. Restricting
ourselves to this lowest transverse channel, the corresponding
Hamiltonian on a finite-difference grid in the $z$-direction may
be written \cite{Rejec_rejec99}:
\begin{eqnarray}
H=-t\sum_{i,\sigma }\left( c_{i+1,\sigma }^{\dagger }c_{i,\sigma
}+\mathrm{h.c.}\right) +\sum_{i}\epsilon _{i}n_{i}+ \label{Rejec_h}
\frac{1}{2}\sum_{i\neq
j}U_{ij}n_{i}n_{j}+\frac{1}{2}\sum_{i,\sigma }U_{ii}n_{i,\sigma
}n_{i,-\sigma }. 
\end{eqnarray}
This is a general form, the difference between different wire
shapes being reflected entirely in the energy parameters
$\epsilon $ and $U$. We note that this Hamiltonian also has the
form for a perfectly straight wire subject to a smooth potential
variation, defined by the $\epsilon $.

In order to study the many-electron problem, it is convenient to
express the Hamiltonian in a basis which distinguishes bound and
unbound states explicitly. This may be done by first diagonalizing the
single-electron part of Eq.~(\ref{Rejec_h}) with the transformation
$c_{q,\sigma }=\sum_{i}\phi _{i}^{q}c_{i,\sigma }$ with eigenenergies
$\varepsilon _{q}$. In this basis the Hamiltonian becomes,
\begin{equation}
H=\sum_{q}\epsilon _{q}n_{q}+\frac{1}{2}\sum\limits_{\scriptstyle
q_1 q_2 q_3 q_4  \atop
  \scriptstyle \sigma \sigma '} {{\cal U}(q_{1}q_{2}q_{3}q_{4})c_{q_{1},\sigma
}^{\dagger }c_{q_{3},\sigma ^{\prime }}^{\dagger }c_{q_{4},\sigma
^{\prime }}c_{q_{2},\sigma }},  \label{Rejec_h2q}
\end{equation}
where ${\cal U}(q_{1}q_{2}q_{3}q_{4})=\sum_{ij}U_{ij}\phi
_{i}^{q_{1}}(\phi _{i}^{q_{2}})^{\ast }\phi _{j}^{q_{3}}(\phi
_{j}^{q_{4}})^{\ast }$. We further denote the lowest localized state with
energy $\epsilon _{q}<0$ by $d_{\sigma }\equiv c_{q,\sigma
},$ with $n_{d}=\sum_{\sigma }d_{\sigma }^{\dagger }d_{\sigma }$
and, similarly, the scattering states with positive $\epsilon _{q}$ are
distinguished by $q\rightarrow k$. There are two independent unbound states
corresponding to each $k$ and these are chosen to be plane waves
asymptotically, i.e. $\phi _{j}^{k}\rightarrow e^{ikj}$ as $j\rightarrow \pm
\infty $ and $\epsilon _{k}=\frac{\hbar ^{2}k^{2}}{2m^{\ast }}.$ Retaining
only those Coulomb matrix elements which involve both localized and
scattered electrons, omitting all terms which would give rise to states in
which the localized state is unoccupied, we arrive at an Anderson-type
Hamiltonian,
\begin{eqnarray}
H &=&\sum_{k}\epsilon _{k}n_{k}+\epsilon _{d}n_{d}+ \sum_{k,\sigma
}(V_{k}n_{d,-\sigma }c_{k,\sigma }^{\dagger }d_{\sigma }+
\mathrm{h.c.})+\\&&+Un_{d\uparrow }n_{d\downarrow } +\sum_{kk^{\prime
},\sigma }M_{kk^{\prime }}n_{d}c_{k,\sigma }^{\dagger }c_{k^{\prime
},\sigma }+\sum_{kk^{\prime } }J_{kk^{\prime }}{\bf S }_{\bf d}\cdot
{\bf s}_{\bf kk^{\prime }}.\label{Rejec_k} \nonumber
\end{eqnarray}
Here $U={\cal U}(dddd)$ is the Hubbard repulsion,
$V_{k}={\cal U} (dddk) $ is mixing term, $M_{kk^{\prime
}}={\cal U}(ddkk^{\prime })-\frac{
1}{2}{\cal U}(dkk^{\prime }d)$ corresponds to scattering of
electrons and the direct exchange coupling is $J_{kk^{\prime
}}=2\,{\cal U}(dkk^{\prime }d)$. Spin operators in
Eq.~(\ref{Rejec_k}) are defined as ${\bf S}_{\bf d}=\frac{1
}{2}\sum_{\sigma \sigma ^{\prime }}d_{\sigma }^{\dagger }{
\hbox{\boldmath$\sigma$}}_{\sigma \sigma ^{\prime }}d_{\sigma
^{\prime }}$ and ${\bf s}_{\bf kk^{\prime
}}=\frac{1}{2}\sum_{\sigma \sigma ^{\prime }}c_{k,\sigma
}^{\dagger }{\hbox{\boldmath$\sigma$}}_{\sigma \sigma ^{\prime
}}c_{k^{\prime },\sigma ^{\prime }}$, where
${\hbox{\boldmath$\sigma$}} _{\sigma \sigma ^{\prime }}$ are the
usual Pauli matrices. Although the Hamiltonian, Eq.~(\ref{Rejec_k}), is
similar to the usual Anderson Hamiltonian~\cite{Rejec_mahan}, we stress
the important difference that the $kd$-hybridisation term above
arises solely from the Coulomb interaction, whereas in the usual
Anderson case it comes primarily from one-electron interactions.
These have been completely eliminated above by solving the
one-electron problem exactly. The resulting hybridization term
contains the factor $n_{-\sigma }$, and hence disappears when the
localized orbital is unoccupied. \ This reflects the fact that an
effective double-barrier structure and resonant bound state
occurs via Coulomb repulsion only because of the presence of a
localized electron.

\begin{figure}[tbh]
\center{\epsfig{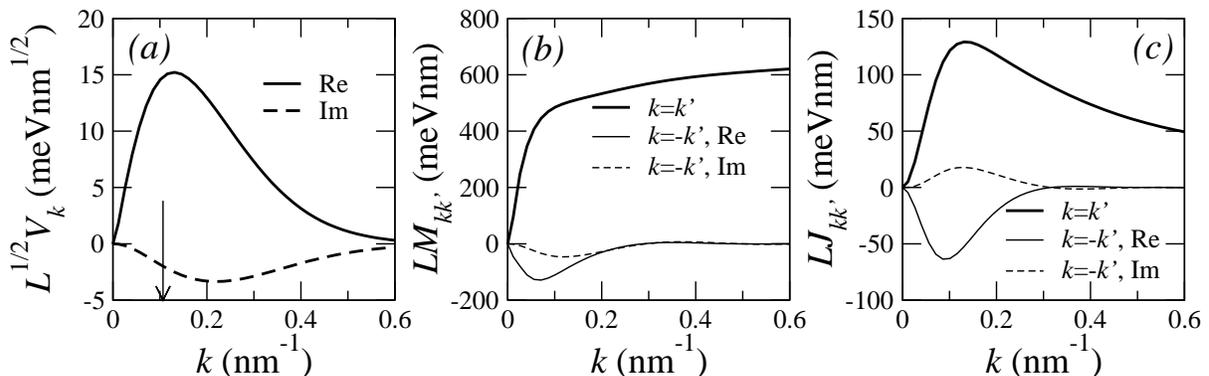}}
\caption{$k$-dependence of matrix elements of the effective Anderson
model. The wire is parametrised as in
Ref.~\protect\cite{Rejec_rejec99} (with $ a_{0}=10$~nm, $\protect\xi
=0.24$, $a_{1}/a_{0}=2$, $V_{0}=0.4$~eV, $\protect \rho=50$~nm and
$\protect\gamma =1$) (a) Mixing coupling $V_{k}$. The energy
$\epsilon_d + U$ is indicated with an arrow. (b, c) Scattering
couplings $M_{kk'}$ and $J_{kk'}$.}
\label{Rejec_Fig1}
\end{figure}

\begin{figure}[tbh]
\center{\epsfig{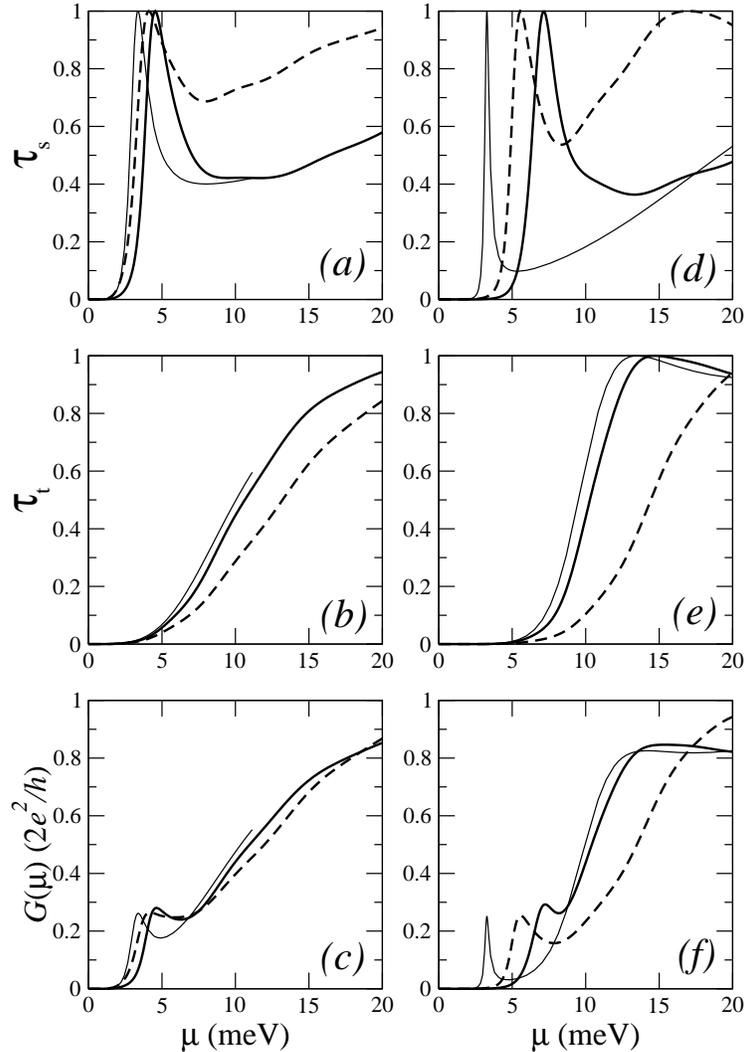}}
\caption{Singlet (a, d) and triplet (b, e) transmission probabilities and
corresponding conductances (c, f). Parameters for the left set are as in
Fig.~\ref{Rejec_Fig1}, for the right set: $a_{0}=10$~nm, $\protect\xi =0.15$,
$a_{1}/a_{0}=4$, $V_{0}=0.4$~eV, $\protect\rho=100$~nm and
$\protect\gamma =0.9$.  Thin lines represent exact results from
Eq.~(\ref{Rejec_h}), thick lines are results from
Eq.~(\ref{Rejec_k}). Dashed lines show results where the exchange
term in Eq.~(\ref{Rejec_k}) is neglected.} \label{Rejec_Fig2}
\end{figure}

In Fig.~\ref{Rejec_Fig1} couplings $V_{k}$, $M_{kk^\prime}$ and
$J_{kk^\prime}$ are shown for a set of wire parameters used in
Fig.~3(a) of Ref.~\cite{Rejec_rejec99}.
The scattering solutions of the Hamiltonian (\ref{Rejec_k}) were
obtained exactly for two electrons with the boundary condition
that for $z\rightarrow \infty $, one electron occupies the lowest
bound state, whilst the other is in a forward propagating plane
wave state, $\phi _{k}(z)\sim e^{ikz}.$ From these solutions we
compute the conductance using the Landauer-B\"{u}ttiker formula
which, incorporating the results of spin-dependent scattering,
takes the following form \cite{Rejec_rejec99},
\begin{equation}
G(\mu )=\frac{2e^{2}}{h}\int \frac{-\partial f(\epsilon -\mu
,T)}{\partial \epsilon }[\frac{1}{4}{\cal T}_{\mathrm{s}}(\epsilon
)+\frac{3}{4} {\cal T}_{\mathrm{t}}(\epsilon )]\mathrm{d}\epsilon
\label{Rejec_conductance}
\end{equation}
where ${\cal T}_{\mathrm{s}}$ and ${\cal T}_{\mathrm{t} } $
are the singlet and triplet transmission probabilities
respectively and $\mu $ is the Fermi energy in the leads. In
Fig.~\ref{Rejec_Fig2} ${\cal T}_{ \mathrm{s}}$,
${\cal T}_{\mathrm{t}}$ and conductance $G(\mu )$ are presented.

The thin lines are the exact scattering result for two electrons.
This shows that quantum wires with weak longitudinal confinement,
or open quantum dots, can give rise to spin-dependent, Coulomb
blockade resonances when a single electron is bound in the
confined region. This is a universal effect in one-dimensional
systems with very weak longitudinal confinement. The emergence of
a specific structure is a consequence of the singlet and triplet
nature of the resonances and the probability ratio 1:3 for
singlet and triplet scattering and as such is a universal effect.
The solid lines show the exact scattering solutions for the
Anderson-type Hamiltonian, for which the matrix elements, and
their energy dependence are calculated explicitly. The solution
of this Anderson-type model for two electrons, in which the
localized level always contains at least one electron, reproduce
the main features of the exact scattering solutions of the
original model. The energy dependence of the matrix elements is
essential to get this good agreement. We have also solved a
similar model in which plane waves, rather than exact scattering
states of the non-interacting problem, were used. However, this
gave poor agreement with the exact results. \ We conclude that an
Anderson-type model is adequate for a near-perfect quantum wire
provided that a suitable basis set is used and the
energy-dependence of the matrix elements is accurately
determined. Future work will focus on the many-electron
properties of this effective Hamiltonian, including `Kondo' and
`mixed valence' regimes.

\begin{acknowledgments}
This work was part-funded by the U.K. Ministry of Defence and the EU.
\end{acknowledgments}


\begin{references}
\bibitem{Rejec_wees88} {B.J. van Wees {\em et al.}, Phys. Rev. Lett. {\bf
60}, 848 (1988); D.A. Wharam {\em et al.}, J. Phys. C {\bf 21}, L209
(1988).}
\bibitem{Rejec_wires} {A. Yacoby {\em et al.}, Phys. Rev. Lett. {\bf 77},
4612 (1996); P. Ramvall {\em et al.}, Appl. Phys. Lett. {\bf 71}, 918
(1997); M. Grundmann {\em et al.}, Semicond. Sci. Tech. {\bf 9}, 1939
(1994); R. Rinaldi {\em et al.}, Phys. Rev. Lett. {\bf 73}, 2899
(1994); K.J. Thomas {\em et al.}, Phys. Rev. Lett. {\bf 77}, 135
(1996); Phys. Rev. B {\bf 58}, 4846 (1998); D. Kaufman {\em et al.},
Phys. Rev. B {\bf 59}, R10433 (1999); D.J. Reilly, cond-mat/0001174.}
\bibitem{Rejec_landauer57}  {R. Landauer, IBM J. Res. Dev. {\bf 1}, 223 (1957);
{\bf 32}, 306 (1988); M. B\"{u}ttiker, Phys. Rev. Lett. {\bf 57},
1761 (1986).}
\bibitem{Rejec_rejec99} {T. Rejec, A. Ram\v sak, and J.H. Jefferson,
cond-mat/9910399 (submitted to Phys. Rev. B);
T. Rejec, A. Ram\v sak, and J.H. Jefferson,
J. Phys.: Condens. Matter {\bf 12}, L233 (2000).}
\bibitem{Rejec_mahan} {G.D. Mahan, {\it Many-Particle Physics}, Plenum
Press, New York (1990).}
\end{references}
\end{document}